\documentclass[conference]{IEEEtran}
\IEEEoverridecommandlockouts
\usepackage{cite}
\usepackage{amsmath,amssymb,amsfonts}
\usepackage{algorithmic}
\usepackage{graphicx}
\usepackage{textcomp}
\usepackage{xcolor}
\usepackage{balance}
\usepackage[italian,english]{babel}
\usepackage{subfigure}

\def\BibTeX{{\rm B\kern-.05em{\sc i\kern-.025em b}\kern-.08em
    T\kern-.1667em\lower.7ex\hbox{E}\kern-.125emX}}
\begin{document}

\title{MQTT-ST: a Spanning Tree Protocol for Distributed MQTT Brokers}

\author{\IEEEauthorblockN{Edoardo Longo\IEEEauthorrefmark{1},
Alessandro E.C. Redondi\IEEEauthorrefmark{1}, Matteo Cesana\IEEEauthorrefmark{1},
Andr\'es Arcia-Moret\IEEEauthorrefmark{2} and Pietro Manzoni\IEEEauthorrefmark{3}}

\IEEEauthorblockA{\IEEEauthorrefmark{1}DEIB, Politecnico di Milano, Italy - Email: \{edoardo.longo, alessandroenrico.redondi, matteo.cesana\}@polimi.it}
\IEEEauthorblockA{\IEEEauthorrefmark{2}Computer Laboratory, University of Cambridge, UK Email: andres.arcia@cl.cam.ac.uk}
\IEEEauthorblockA{\IEEEauthorrefmark{3}DISCA, Universitat Poltècnica de València, Spain - Email: pmanzoni@disca.upv.es}
}

\maketitle

\begin{abstract}
	MQTT, one of the most popular protocols for the IoT, works according to a publish/subscribe pattern in which multiple clients connect to a single broker, generally hosted in the cloud. However, such a centralised approach does not scale well considering the massive numbers of IoT devices forecasted in the next future, thus calling for distributed solutions in which multiple brokers cooperate together. Indeed, distributed brokers can be moved from traditional cloud-based infrastructure to the edge of the network (as it is envisioned by the upcoming MEC technology of 5G cellular networks), with clear improvements in terms of latency, for example. This paper proposes MQTT-ST, a protocol able to create such a distributed architecture of brokers, organized through a spanning tree. The protocol uses in-band signalling (i.e., reuses MQTT primitives for the control messages) and allows for full message replication among brokers, as well as robustness against failures. We tested MQTT-ST in different experimental scenarios and we released it as open-source project to allow for reproducible research.
	
\end{abstract}


\section{Introduction}
\label{sec:introduction}
The advent of the 5th Generation (5G) of mobile cellular networks will boost the development and implementation of large scale, city-wide IoT applications. Indeed, two main 5G innovation pillars have been designed precisely for accommodating IoT requirements: massive Machine Type Communication (mMTC) and Multi-Access Edge Computing (MEC). On the one hand, mMTC will enable connection densities in the order of 10$^6$ low-power devices per square kilometre, with enormous implications on the amount of traffic generated and transmitted. On the other hand, MEC technology will bring computational power, storage resources and service infrastructures to the edge, alleviating the resources needed in the core network and reducing latency.

Such a dramatic change will also have a great impact on the communication protocols associated with the IoT ecosystem. While at the lower layers of the stack the plethora of short-range low-power protocols (e.g., the IEEE802.15.4 family) 
will have to compete with 5G-based solutions for survival, at the application layer the MQTT (Message Queuing Telemetry Transport) protocol is living its greatest period of popularity since its introduction and it can be considered the de-facto standard for IoT solutions\footnote{Not by chance, the four major cloud computing services up to date (Amazon AWS, Google Cloud Platform, IBM Cloud and Microsoft Azure) all adopt MQTT as protocol for connecting IoT devices to their endpoints.}.

\begin{figure}[t!]
  \centering
		\subfigure[]{\label{fig:cloud}\includegraphics[width=0.48\columnwidth]{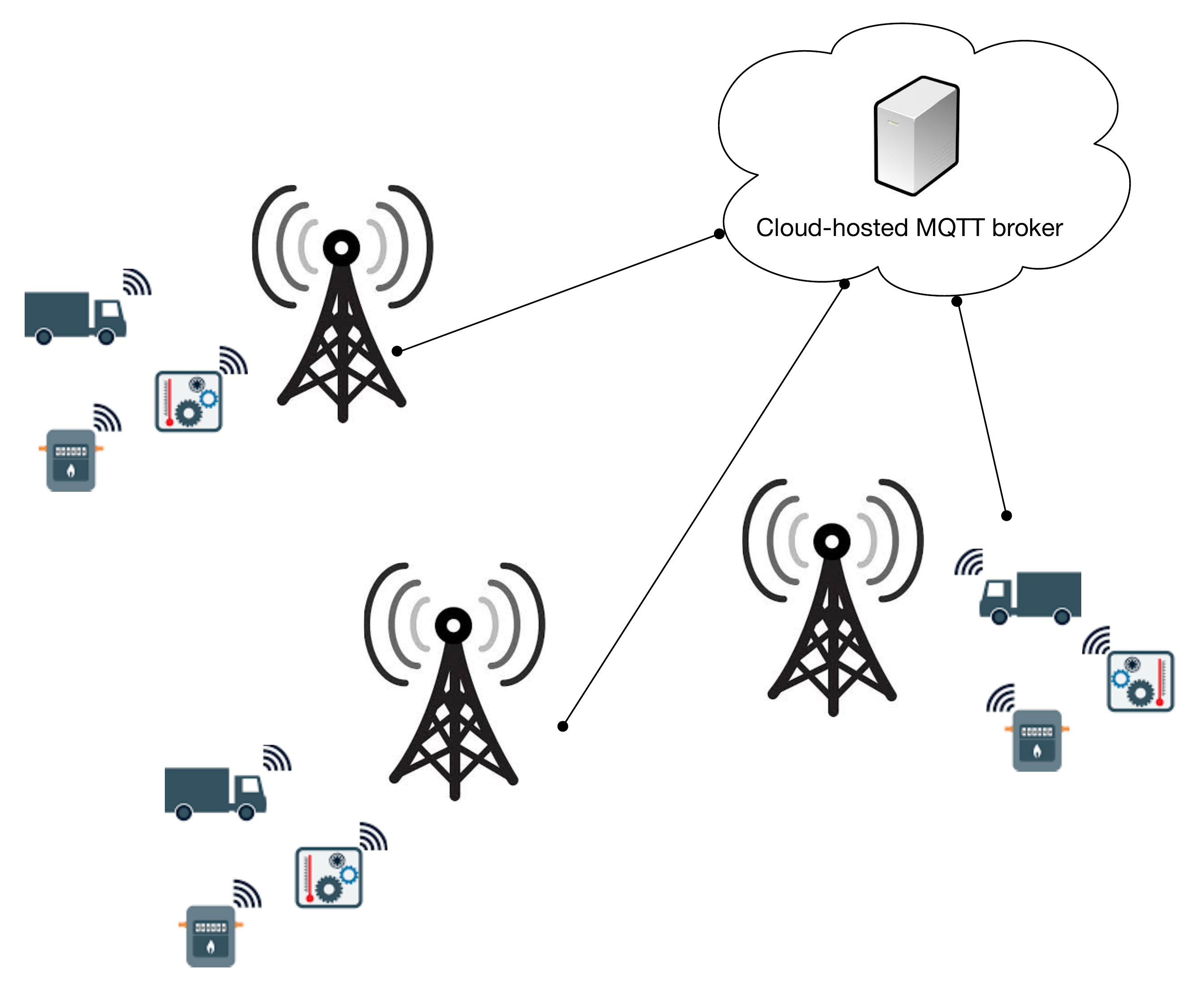}}
		\subfigure[]{\label{fig:edge}\includegraphics[width=0.48\columnwidth]{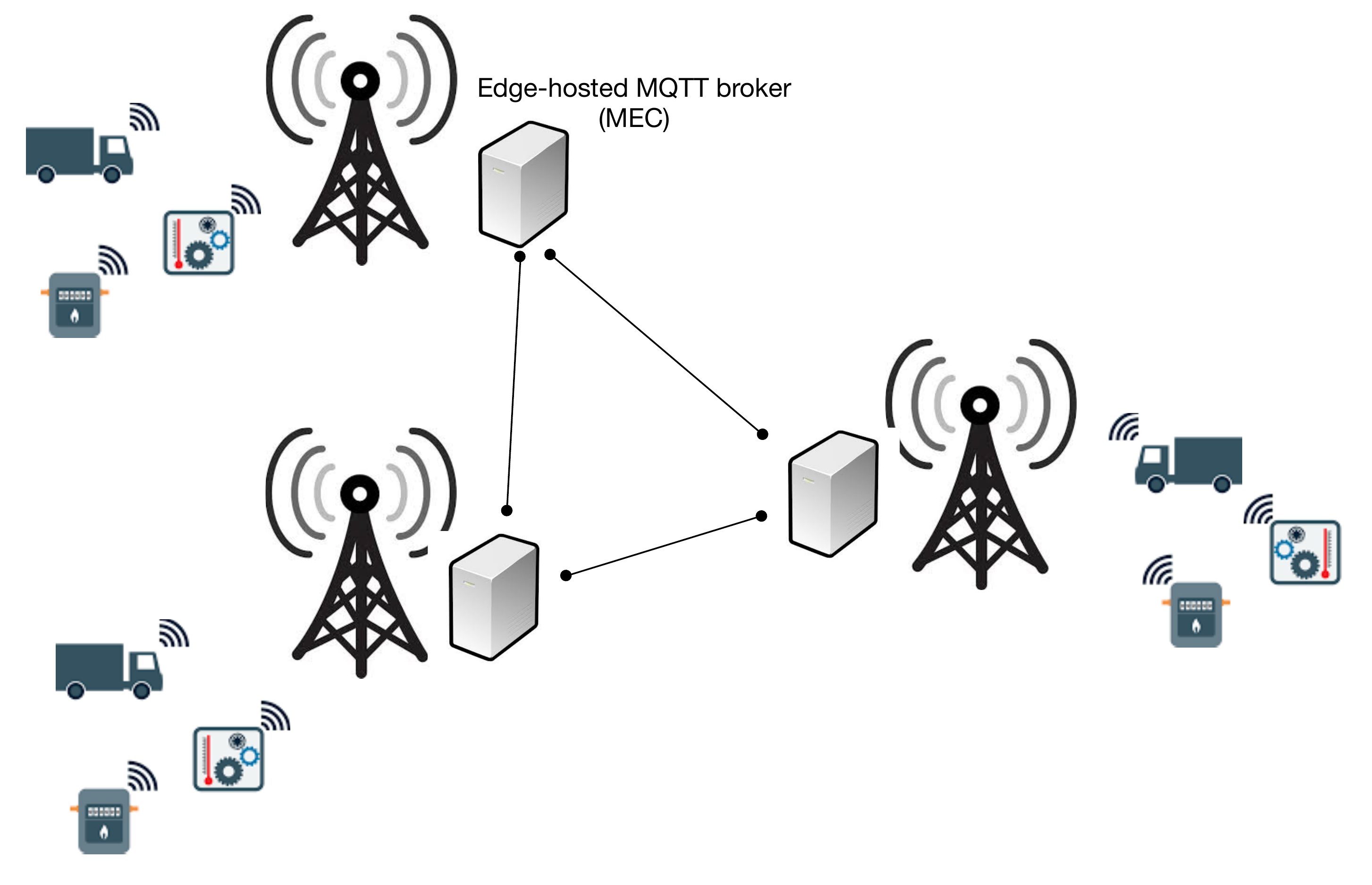}}
  \caption{Centralised broker scenario (a) and distributed architecture (b).}
  \label{fig:edge-cloud}
\end{figure}

MQTT is a lightweight publish/subscribe protocol designed around a central \textit{broker}. Clients connect to the broker and subscribe to or publish data on specific \textit{topics}. The broker is in charge of forwarding the data published to the clients interested in it, thus decoupling the process of data generation and consumption both in space and in time. This aspect, combined with the protocol simplicity at the client side and the support for reliability and quality of service (QoS), makes MQTT ideal for resource-constrained applications and motivates its great popularity. However, MQTT still remains a centralised protocol which nicely fits classical cloud-based architectures, in which all IoT devices connect to a single broker endpoint. 
This picture is partially at odds with the one envisioned by 5G, in which cloud services (including any broker instance) are moved to the edge, closer to the user devices. For MQTT this means moving from the current centralised, star-shaped, single-broker topology to a distributed, multi-broker topology which can cope with the massive numbers of devices envisioned to be served (see Figure \ref{fig:edge-cloud}).

In our previous position paper~\cite{Redondi2019towards} we analysed the main research challenges and possible solutions to scale up a pub/sub architecture for upcoming 5G networks, and we presented our view on system design and optimisation. In this paper we tackle the problem from an implementation perspective: we observe that due to the intrinsic nature of the pub/sub pattern, bridging brokers may result in potentially harmful message loops. Existing solutions solve the problem by imposing a static tree-based topology among brokers, which lacks adaptivity and robustness. Therefore, in this paper we propose and implement MQTT-ST, a protocol inspired by STP (Spanning Tree Protocol) for interconnecting MQTT brokers automatically in a loop-free topology, in order to
distribute messages among them. The protocol uses in-band signalling (i.e., all control messages are embedded in MQTT native mechanisms) and allows to have full message replication as well as robustness against node failures. We test MQTT-ST in different experimental scenarios, focusing on latency, computational load and achievable throughput and we compare it with traditional cloud-based, single-broker approaches. Finally, we release MQTT-ST as open-source project based on the popular Eclipse Mosquitto MQTT broker, in order to allow for reproducible research.




\section{System Overview}
\label{sec:system_overview}


\subsection{Motivation}
Some existing MQTT broker implementations (e.g. Mosquitto, CloudMQTT, HiveMQ) allow the use of bridging, i.e., a direct connection between brokers. The feature allows a broker \texttt{B} to connect to another broker \texttt{A} as a standard client, subscribing to all or a subset of the topics published by clients to \texttt{A}. Unfortunately, such a procedure is prone to message loops among brokers: indeed, the existence of a cycle where a message is continuously republished by the participating brokers, can quickly deplete a broker's resources, ultimately making it unable to deliver meaningful traffic. Due to the enormous complexity of implementing duplicate detection in distributed scenarios, which would require to keep track of the original producer of every message received and forwarded by any broker, existing solutions require to manually configure the connections between brokers in a loop-free topology, i.e., a tree. However, such a manual configuration of MQTT bridges has two main drawbacks: first, similarly to wiring switches in small- or medium-size enterprises, it can become a very confusing task with a high chance of creating accidental duplicate connections, especially in large topologies. Second, by enforcing a loop-free static topology among brokers, adaptivity and robustness to failures are completely lost.
Another option is to rely on an automatic way to set up a tree among brokers. In switched networks, this is obtained using the Spanning Tree Protocol (STP) and/or its following amendments: the purpose of this paper is to embed STP mechanisms into MQTT, ultimately allowing for the creation of a loop-free, dynamic and robust network of brokers.

\subsection{Spanning Tree Protocol}
The Spanning Tree Protocol (standardised as IEEE 802.1D) is a distributed protocol which creates a logical spanning tree over a meshed network of layer 2 switches. Such a spanning tree is obtained by electing a root switch and blocking some of the output ports of the other switches: blocked ports do not forward data frames, thus avoiding broadcast storms. In order to agree on the root node and on which ports should be blocked, switches exchange control packets known as BPDU (Bridge Protocol Data Unit). 

In a nutshell, the main steps of STP are the following:
\begin{itemize}
	\item{At startup, each node sets itself as root and start broadcasting BPDU. Each BPDU contains among other parameters) the identifier of the node and the transmitting port, the identifier of the current root node selected by the transmitting node and the root patch cost. The node identifier, composed of both the node MAC address and a configurable priority value is used for root selection: the node with the lowest identifier is elected as root.}
	\item{Upon the reception of a BPDU, a node reconfigures its state by modifying the identifier of the (believed) root node and updating the root port (port that leads to the least cost path to the root). The rest of the active ports are labeled as either designated (used for forwarding traffic) or blocked. To avoid loops, nodes agree on which port should be designated or blocked, based again on the least-cost path to the root or the lowest identifier, in case of ties.}
	\item{BPDUs are periodically transmitted by the root, and forwarded by all other nodes, in order to keep the topology updated. Upon failures on active links, special BPDU known as Topology Change Notification (TCN) can be transmitted by non-root nodes to inform all the others.}
\end{itemize}
Enhancements of this general scheme were proposed and standardised in several standards, including the Rapid Spanning Tree Protocol (RSTP, IEEE802.1D-2004), which provides significantly faster spanning tree convergence, the Multiple Spanning Tree Protocol (MSTP, IEEE802.1Q-2005), which allows the creation of multiple spanning trees for different VLANs or group of VLANs and recently the Shortest Path Bridging protocol (IEEE 802.1aq), which allows redundant links between switches to be active at the same time, thus increasing bandwidth.


\subsection{MQTT-ST}
We develop MQTT-ST starting from the latest MQTT protocol specifications~\cite{MQTT5}. The main changes and modifications to the protocol are reported below.

\subsubsection{Connection phase}
At startup, a broker willing to create a bridge with another broker transmits a MQTT CONNECT message. The address and port of the broker (or brokers) to connect to is specified in a configuration file. To inform a broker that the connection request comes from another broker and not from a client, we set the most significant bit of the Protocol Version byte in the CONNECT header. Upon the reception of a CONNECT message with such a bit set, a broker performs the following operations:
\begin{itemize}
	\item First, to allow bidirectional communication, the broker transmit back a modified CONNECT message to the originating node, using the standard MQTT port 1883. Note also that this allows a node with no configuration file set to be part of the broker network, if contacted by an already connected broker. 
	\item The broker stores the IP addresses of all directly connected brokers in a local table, which is used to keep track of the state of each connection marked as root, designated or blocked. For each connection, the table also stores the average Round Trip Time (RTT) and a value $C$, which summarises the resource capability of the endpoint broker as detailed later.
	\item Finally, the broker sets itself as root and start transmitting signalling messages towards all connected brokers. Instead of creating a new specific message, we reuse MQTT PINGREQ messages.
\end{itemize}

\subsubsection{Signalling phase}
Standard MQTT specifies a Keep Alive parameter, which defines the maximum time interval permitted to elapse after the last client transmission. In case the timer expires, the broker closes the connection with the client. Therefore, to maintain the connection alive, a client transmits periodical PINGREQ messages. MQTT-ST reuses such messages, which play the role of STP BPDUs. In details, the following information is appended to PINGREQ messages: IP address of the current root broker for the transmitter, the broker capability value $C$, and a root path cost $P$. The latter two fields are used for root selection and path computation, respectively.

\subsubsection{Root selection}
The root broker plays a crucial role in the broker tree, as it is the relay node for all traffic and it is therefore subject to an increased computational load. Indeed, selecting a broker with poor or overloaded resources may result in poor overall performance. In STP the root is selected based only on its identifier, which does not suit well the scenario under consideration. In MQTT-ST, instead, the root broker is selected according to the capability value $C$, defined as:
\begin{equation} 
	C = \alpha L + \beta M
\end{equation}
where $L$ is the broker CPU speed, $R$ is the amount of RAM memory and $\alpha$, $\beta$ are tuneable conversion parameters\footnote{Operatively, $L$ and $R$ are read by the broker at startup using the \texttt{/proc/cpuinfo} and \texttt{/proc/meminfo} system files available on Linux}.
In case of tie, the broker with the lowest IP address is selected as root. 


\subsubsection{Path computation}
In STP, each node selects the best path to the root according to a bandwidth-related criteria, in order to avoid the use of reduced capacity links in the tree which may slow down an entire network. For MQTT-ST we observe that latency, rather than bandwidth, plays a critical role. Each broker therefore continuously monitor the RTT to other brokers and uses that value for updating the root path cost $P$. In order to do this, we leverage the request/response mechanism already present in MQTT through the PINGREQ/PINGRESP. A timer is started when the client transmits the PINGREQ message and it is stopped when the corresponding PINGRESP message is received by the broker, providing an estimate of the current RTT.
Upon reception of a PINGREQ message, the connection providing to the lowest latency path towards the root is marked as root connection. In case of ties in the cumulative latency to the root, the connection passing through the broker with the highest amount of resources is selected as root connection. All the other connections are labeled following the same logic of the STP protocol: (i) all connections of the root broker as marked as designated, (ii) the non-root connections of other brokers are marked as designated if the broker as a better path cost (or a better value of $C$ in case of tie) compared to their neighbouring broker and (iii) all other ports are labeled as blocked.

\subsubsection{Runtime behaviour} At runtime, a MQTT-SN broker works exactly like a MQTT broker from the perspective of the connected clients. Moreover, for every published message on any topic, the broker forwards it on its non-blocked connections (root and designated), while any message incoming on a blocked connection is discarded. The forwarding is performed like a standard MQTT PUBLISH message. We highlight that:
\begin{itemize}
	\item To allow full replication of the messages published at one broker to all other brokers, message forwarding is performed with the highest MQTT QoS available (QoS = 2). On the one hand, this guarantees that each message is received only once by the intended recipients, while on the other hand it requires a four-part handshake with a non-negligible associated overhead.

	\item Since forwarding is implemented through a standard MQTT PUBLISH, all other features of the latest MQTT specification are conserved (e.g., retain, topic alias, message expiry interval, etc.)
	\item MQTT-ST also forwards Last Will and Testament (LWT) messages, which are automatically generated by a broker upon ungraceful disconnection of a client. 
	
\end{itemize}

\subsubsection{Reaction to failures}
Upon a broker failure, MQTT-SN handles the corresponding socket error to re-establish the forwarding tree. In details, the broker detecting the socket error transmits a special PINGREQ message used to restart the tree construction from scratch. The broker sets itself as root and append to the message an additional Topology Change (TC) field set, similarly to what happens in STP. Any broker receiving such a message restart the root selection procedure, which eventually will converge to a new tree.

\section{Experimental results}
\label{sec:experimental_results}

MQTT-ST has been developed in C language starting from the open-source project Eclipse Mosquitto~\cite{light2017mosquitto}, which offers both broker and client capabilities. In order to test its functionality and evaluate its performance, two simulation environments have been created.

\subsection{Local environment}\label{sec:le}
The first simulation environment is aimed at performing stress tests on the processes implementing the brokers without looking at network-related factors (e.g. link delay or link capacity). For this reason, the stress test simulation consists in a script which automatically creates a given number of MQTT-ST broker processes, all running on the same machine (Intel i7-3770 with 8 CPUs @ 3.40GHz with 16 GB of RAM, running Ubuntu 16.04.6) and connected together in a fully meshed network, i.e., each broker has connections to all other brokers. To generate client publications and subscriptions, we use a simple open-source benchmark tool\footnote{https://github.com/krylovsk/mqtt-benchmark}.
Such a tool, which is executed on a different machine in the same private LAN of the one running the brokers, generates a given number of clients (both publishers and subscribers), allowing to tune several parameters (e.g., messages size, number of topics, etc.) and to measure the associated broker performance. 

\subsubsection{Signalling overhead}
\begin{figure}[t!]
  \centering
	\includegraphics[width=\columnwidth]{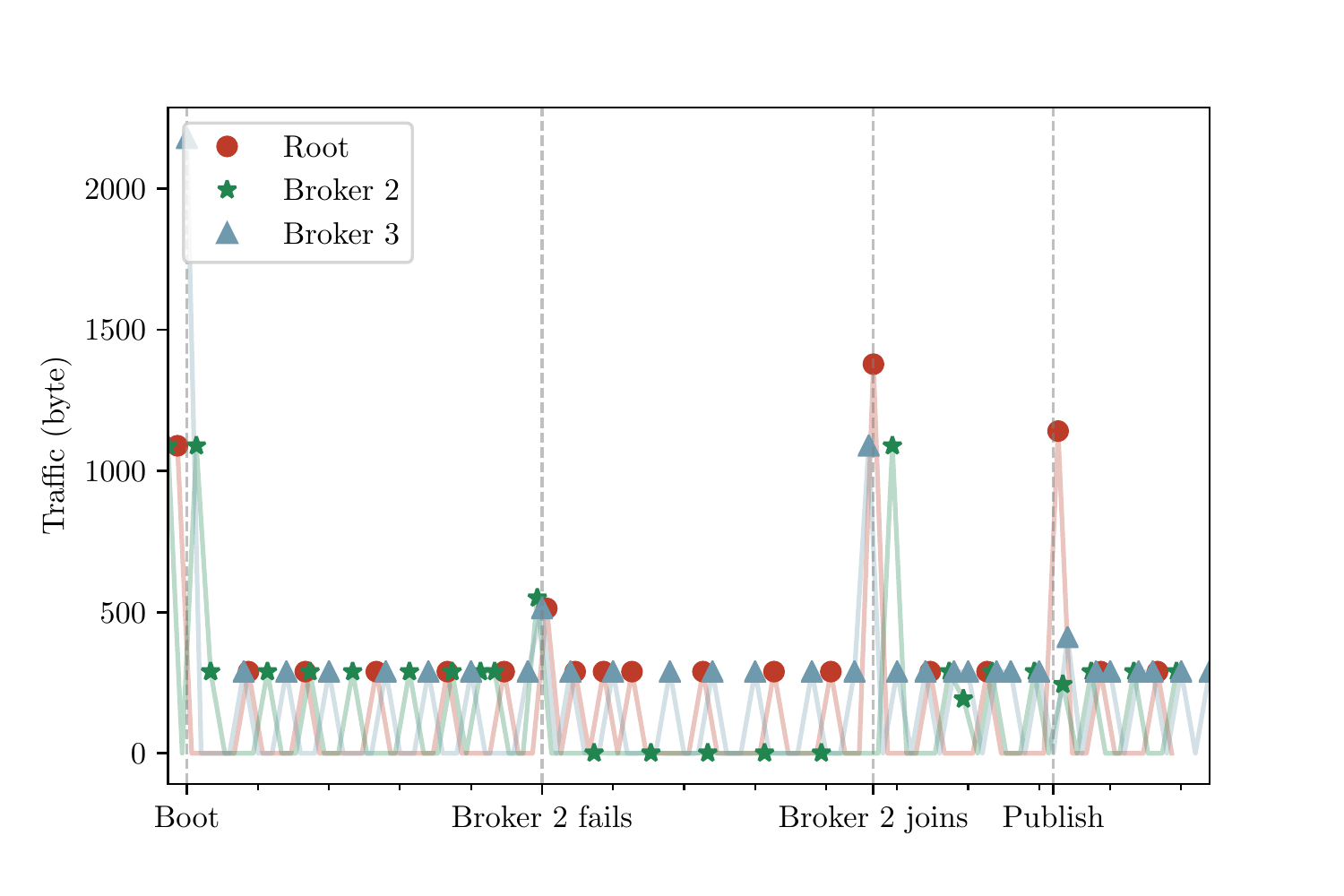}
  \caption{Traffic trace relative to different phases of MQTT-ST operating on three brokers.}
  \label{fig:traffic}
\end{figure}

MQTT-ST requires brokers to exchange information periodically, therefore consuming additional resources compared to the standard MQTT. As explained in section \ref{sec:system_overview}, we modified the PINGREQ messages in order to add BPDU information to them. This adaptation increases the size of the PINGREQ message from 78 to 114 bytes, which is acceptable especially compared to the massive amount of data traffic expected in common IoT scenarios.  
To better show the resource utilisation, we display in Figure~\ref{fig:traffic} the overall traffic (in bytes) consumed by three MQTT-ST brokers when the Keep Alive value is set to 10s; the red line represents the root broker while the other two lines corresponds to brokers connected to the root. We highlight the start up event (Boot), the fail and the subsequent restore of the Broker 2 which triggers the execution of the STP algorithm. Finally, we show a publish event from a client to the root broker, which in turns forwards the publication to all other brokers.

\subsubsection{Publication throughput}\label{sec:pubthr}
A performance measure generally used in related works~\cite{banno2017dissemination, jutadhamakorn2017scalable} is the publication throughput, which is the maximum speed at which a client can publish messages to a broker when using publication QoS = 2 (that is, waiting for the MQTT four way handshake acknowledgment to take place). Since the transmission of a new message needs to wait for the ACK coming from the broker, such a performance measure gives an indication of the current broker workload. We evaluate such a measure in three different scenarios:
\begin{itemize}
	\item Benchmark: $N$ publishers and $M$ subscribers connected to a single, centralised standard MQTT broker. This corresponds to a traditional, cloud-based scenario.
	\item Distributed: $N$ publishers and $M$ subscribers evenly connected to $K$ MQTT-ST brokers (i.e., each broker is connected to $M/K$ publishers and $N/K$ subscribers). This scenario corresponds to the envisioned distributed architecture.
	\item 100\% Locality: $N$ publishers and $M$ subscribers connected to only one of the $K$ MQTT-ST brokers. This is an extreme unfavourable scenario in which messages are distributed to all brokers, but consumed at the same broker where they are produced. 
\end{itemize}
The lefthand side pictures in Figures~\ref{fig:10pub},~\ref{fig:100pub} and~ß\ref{fig:1000pub} show the publication throughput of the three different broker configurations (where the green curve is the average across the $K$ distributed brokers), at different number of subscribers and for 10, 100 and 1000 publishers, respectively. As one can see, the publication throughput of the benchmark scenario rapidly decreases as the number of subscribers and publishers increases, while MQTT-ST always shows higher throughput. The 100\% locality case shows always the worst performance, as it corresponds to the case where one single broker manages all the clients, in addition to forwarding their messages to all other brokers. 

\subsubsection{End-to-end delay}
\begin{figure*}[t!]
   \centering
 		\subfigure[]{\includegraphics[width=0.8\columnwidth]{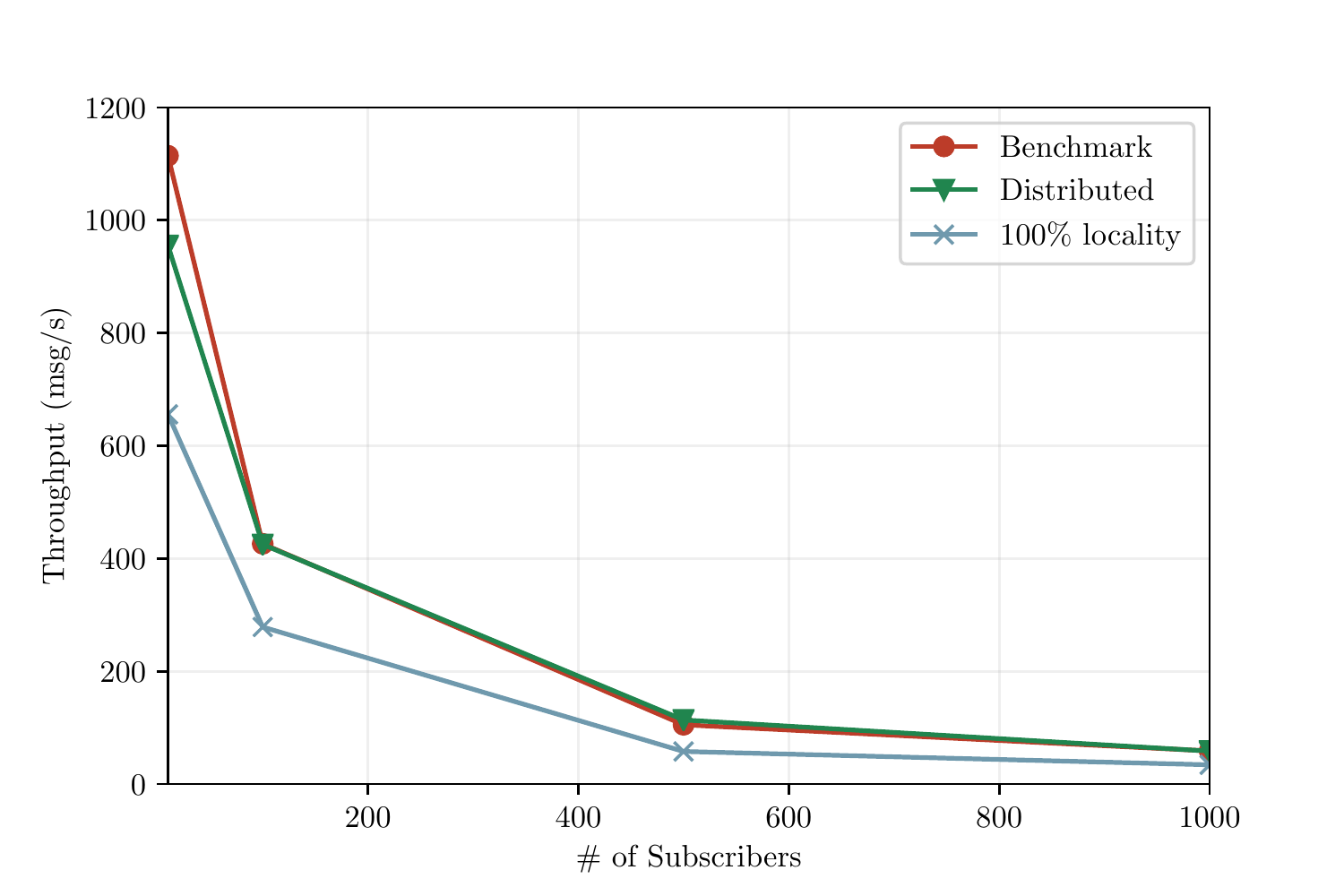}}
 		\subfigure[]{\includegraphics[width=0.8\columnwidth]{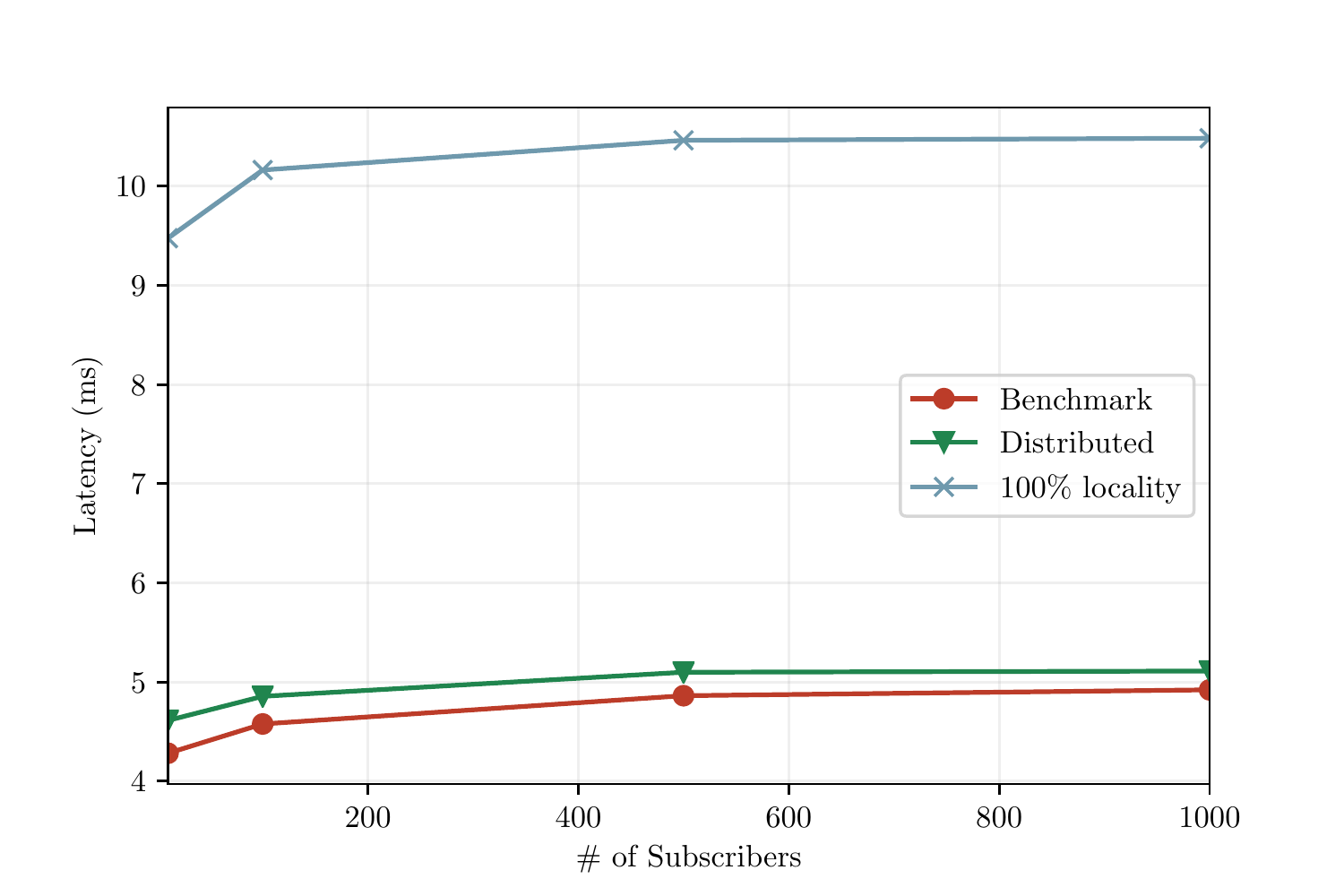}}
   \caption{Publication throughput (a) and end-to-end delay (b) with 10 publishers.} 
   \label{fig:10pub}
 \end{figure*}

\begin{figure*}[t!]
   \centering
 		\subfigure[]{\includegraphics[width=0.8\columnwidth]{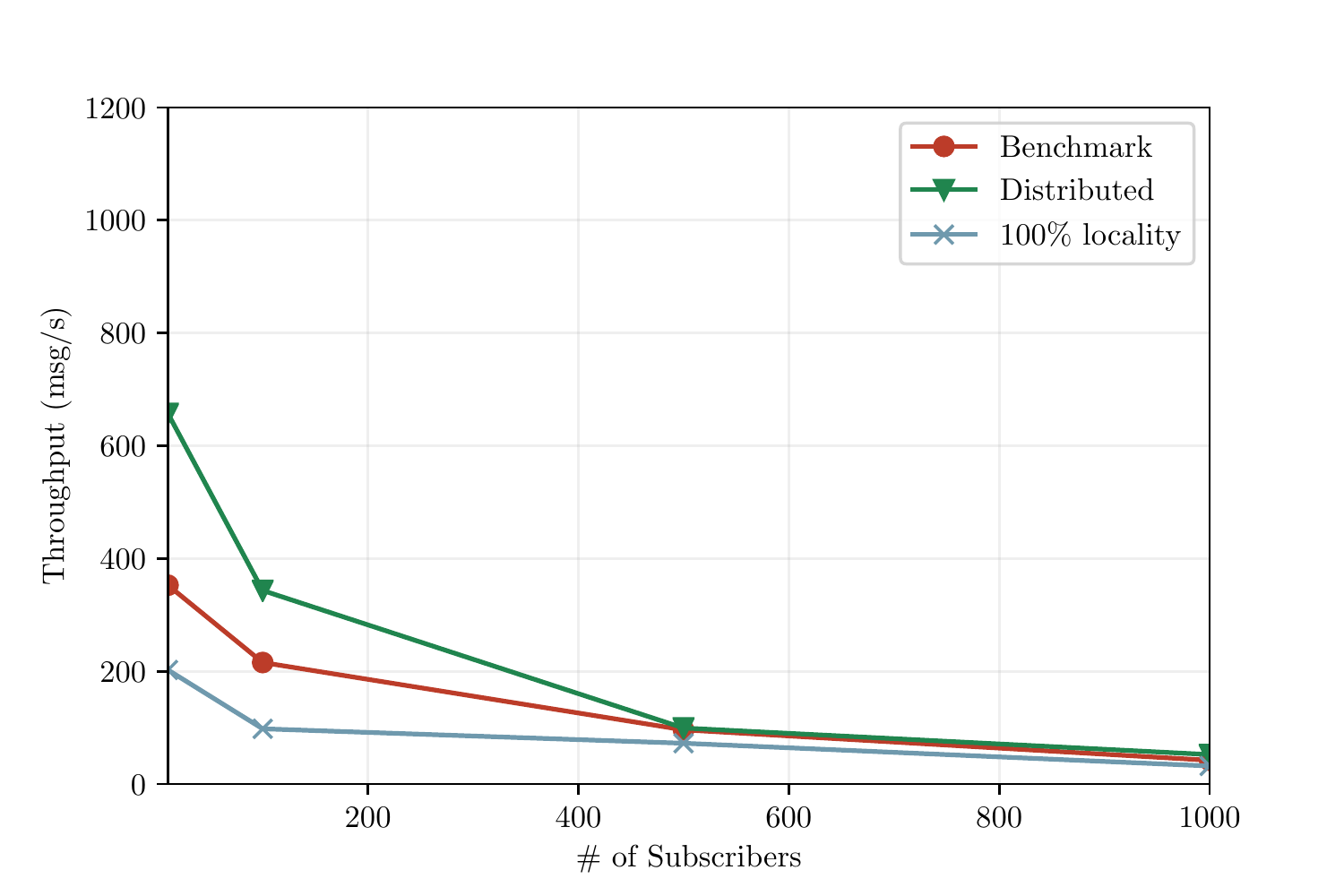}}
 		\subfigure[]{\includegraphics[width=0.8\columnwidth]{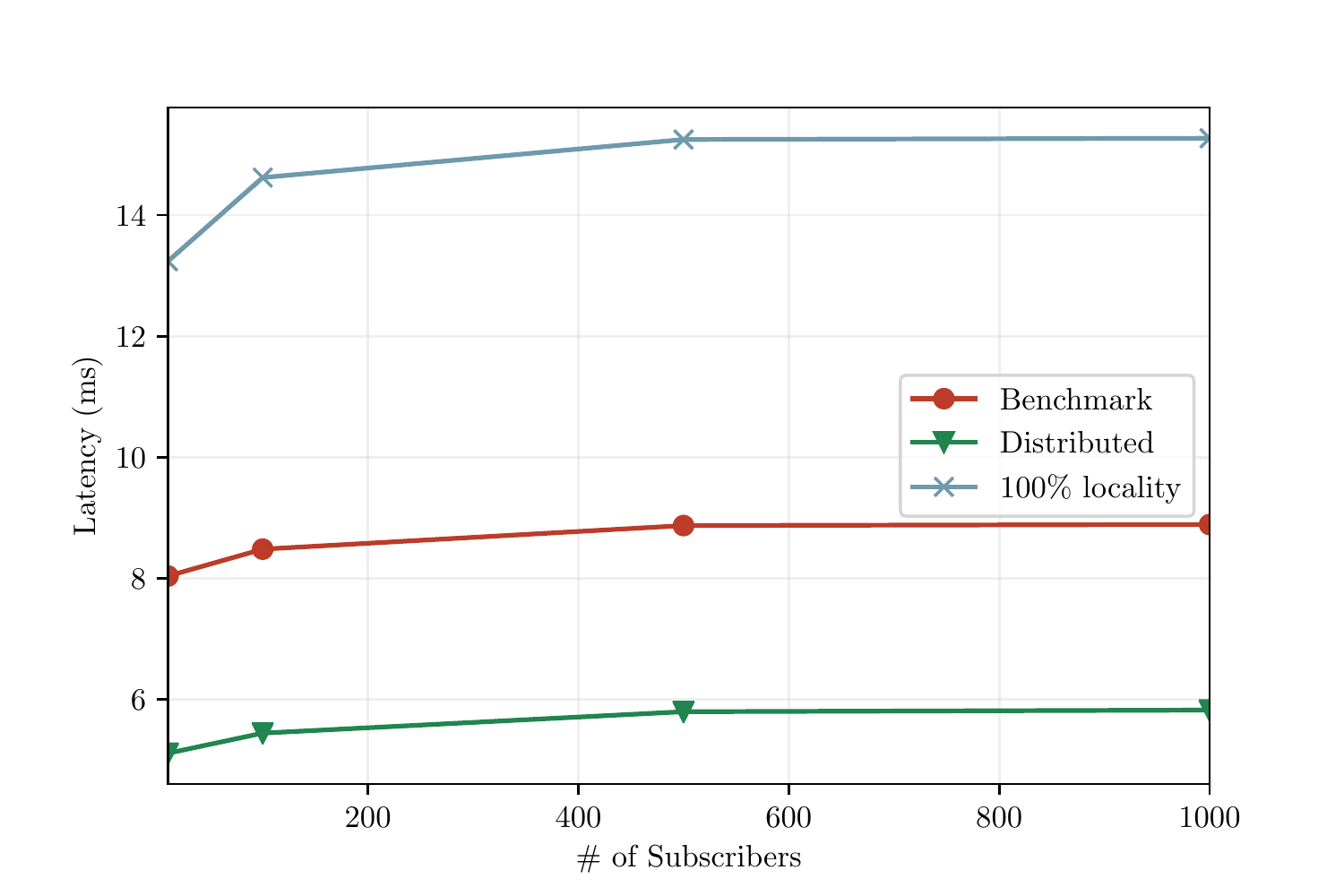}}
   \caption{Publication throughput (a) and end-to-end delay (b) with 100 publishers.}
   \label{fig:100pub}
 \end{figure*}

\begin{figure*}[t!]
   \centering
 		\subfigure[]{\includegraphics[width=0.8\columnwidth]{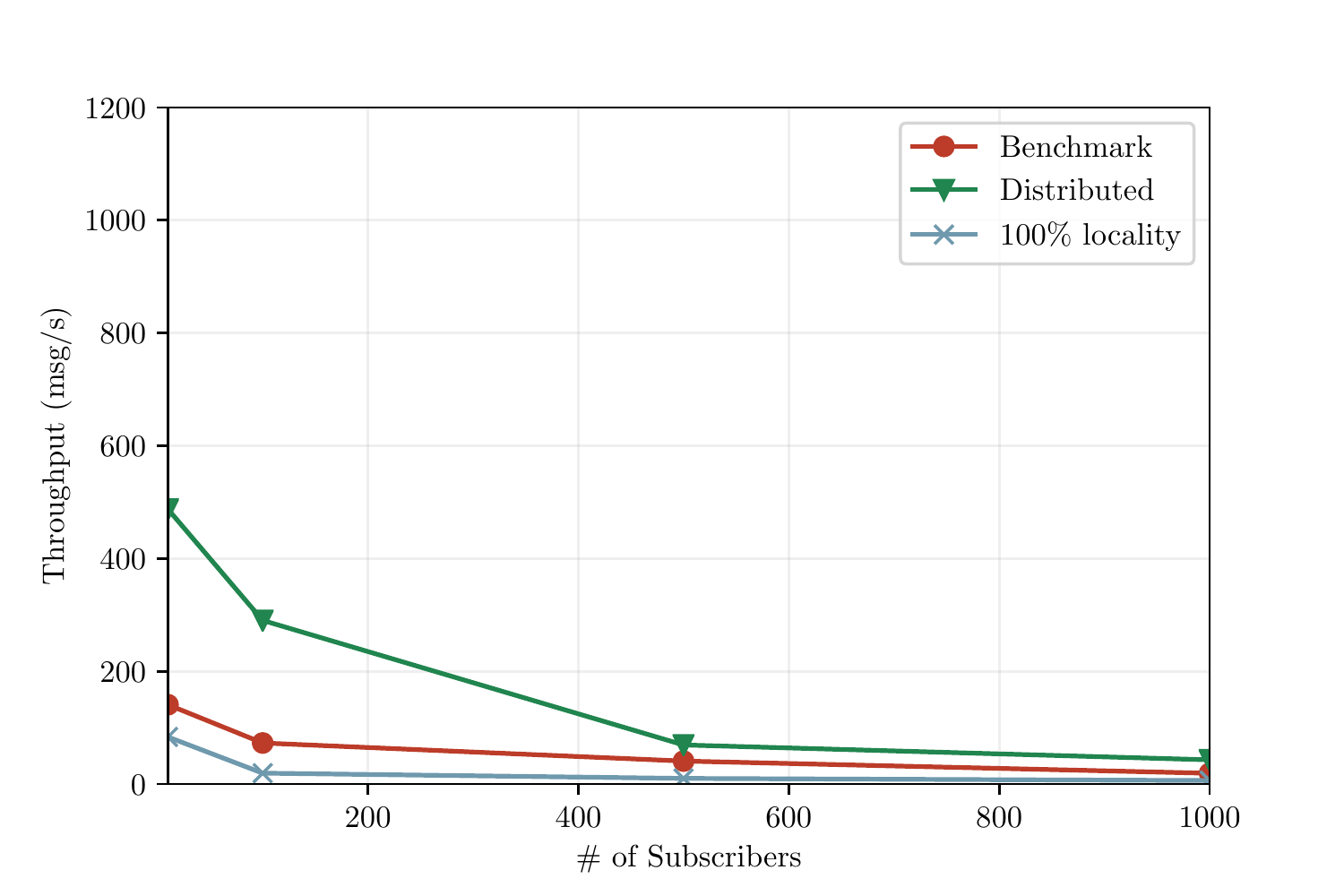}}
 		\subfigure[]{\includegraphics[width=0.8\columnwidth]{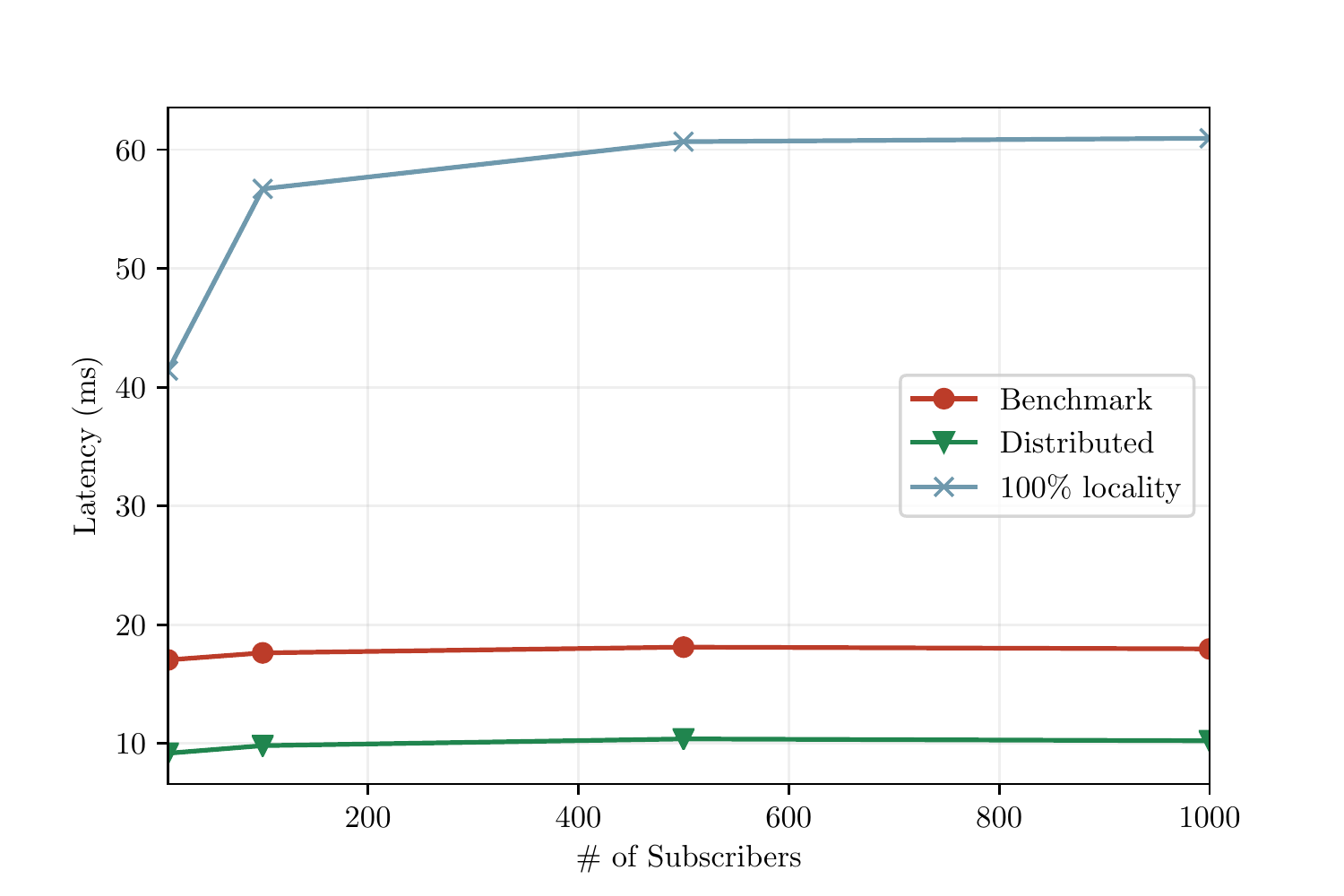}}
   \caption{Publication throughput (a) and end-to-end delay (b) with 1000 publishers.}
   \label{fig:1000pub}
 \end{figure*}

We also evaluate the end-to-end delay (i.e., the average amount of time elapsed between the publication of a message and its reception at a subscriber) in the same aforementioned scenarios, which is reported in the righthand side of Figures  ~\ref{fig:10pub},~\ref{fig:100pub} and~ß\ref{fig:1000pub}. We can observe that (i) the end-to-end delay is sensible to the number of publishers much more than the number of subscribers; (ii) the 100\% locality scenario always shows the worst performance and (iii) MQTT-ST outperforms the benchmark scenario when the number of publishers increases. 


\subsection{Networked environment}
\begin{figure}[t!]
  \centering
		\includegraphics[width=\columnwidth]{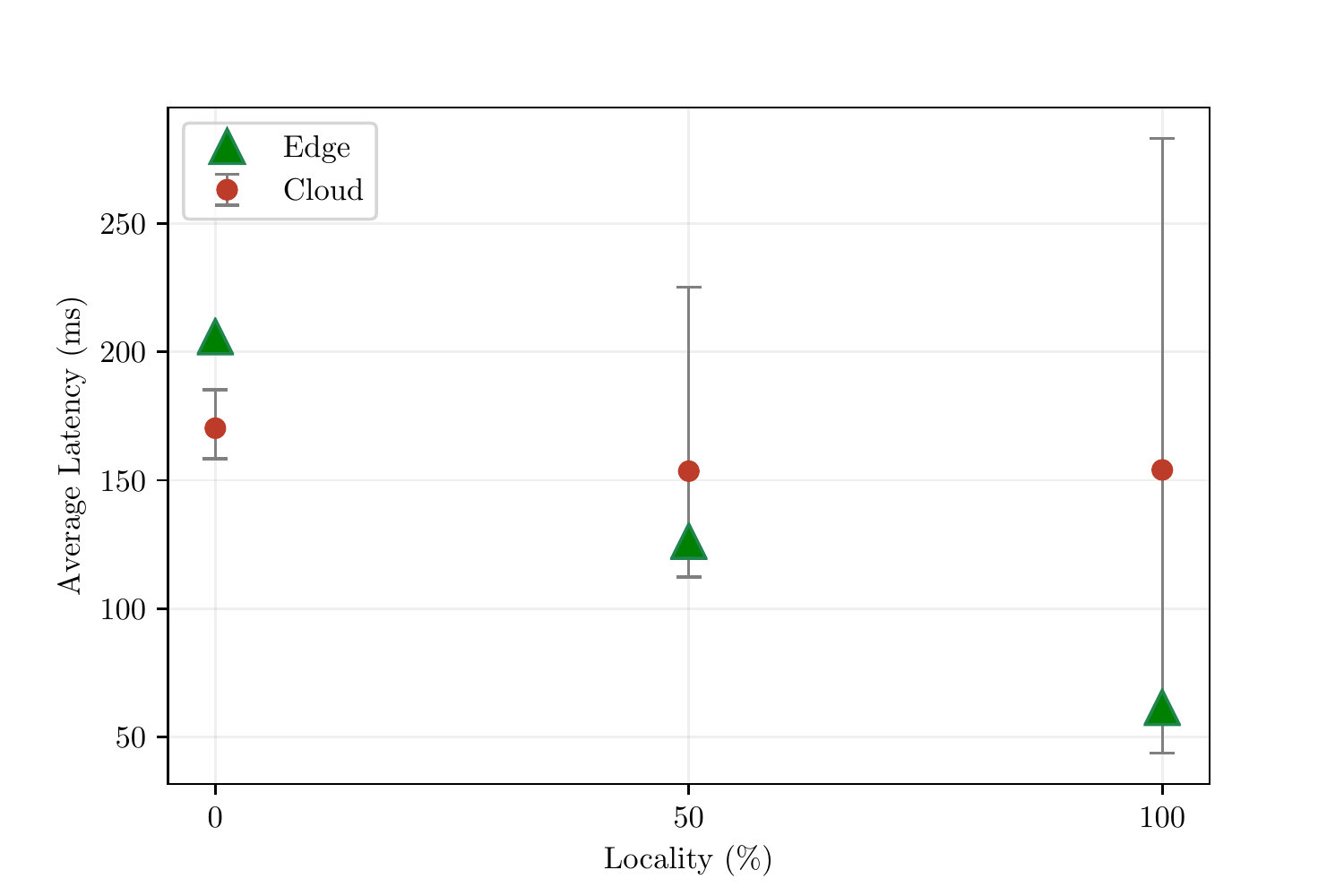}
  \caption{End-to-End latency in a networked scenario.}
  \label{fig:aws}
\end{figure}

The second simulation environment is aimed at demonstrating the functionality of MQTT-ST in a networked scenario compared to a centralised cloud architecture. To this end, we leverage the capabilities of Amazon Web Services EC2 to create several brokers which run on virtual machines (VM) located in different regions of the world. We focus on a scenario with 3 brokers executed on t2.micro VM instances (1 vCPUs @ 2.40GHz, 1 GB RAM) and deployed in the following locations: US West (Oregon), US East (N. Virginia), EU (Ireland). When running the centralised scenario, only one of the brokers is chosen, and the other two are shut off. Clients and subscribers are run on two separate VMs, located in US West (N. California) and EU (Germany). For both the centralised and distributed scenarios, the following tests are performed:
\begin{itemize}
	\item {0\% locality: 100 publishers in N. California and 100 subscribers in Germany}
	\item {50\% locality: 50 publishers and 50 subscribers in both N. California and Germany}
	\item {100\% locality: 100 publishers and 100 subscribers in N. California}
\end{itemize}
Figure~\ref{fig:aws} shows the average end-to-end latency, where the cloud scenario is averaged over all possible locations of the centralised broker. As one can see, MQTT-ST allows to obtain significant latency improvement compared to centralised solutions when publishers and subscribers are colocated. Moreover, the additional cost for full message replication in case publishers and subscribers are far from each other (0\% locality) is limited to few tens of milliseconds.

\section{Related Work}
\label{sec:related_work}

The research area of message dissemination in distributed generic pub/sub system has been very active in the last 20 years. Most works focus on the development of efficient and scalable routing algorithms to create topic-based dissemination trees (in the form of multicast groups) that cover only the subscribers matching a particular topic~\cite{Baldoni2007SIENA,majumder2009scalable,martins2010routing,siegemund2015self,turau2017scalable}. In such works, no specific broker implementation is considered and the overlay broker topology is assumed to be known.
Only very recently, motivated by the protocol popularity, some attention has been given to the problem of interconnecting  MQTT-specific brokers~\cite{horizontalIoT}. Some works focused primarily on vertical clustering, where the single broker is replaced by many virtualized broker instances running behind a single end point, typically a load balancer~\cite{jutadhamakorn2017scalable}\cite{sen2018highly}. These approaches introduce the concept of multiple brokers cooperating with each other, although the broker cluster is seen as a single centralized entity from the perspective of clients. Pure MQTT broker distribution is introduced in Banno et al. in~\cite{banno2017dissemination}: authors propose ILDM (Internetworking Layer for distributed MQTT brokers), where heterogeneous brokers are connected with each other through specific nodes, placed between clients and brokers. Similar to our work, message distribution is obtained with publication flooding, but the underlying network of ILDM nodes is assumed to be already loop-free. Also no automatic mechanisms for broker failure recovery are present. In~\cite{schmitt2018dynamic} and~\cite{schmitt2019data} authors also propose to interconnect MQTT brokers, with the possibility of dynamically changing the topology configuration at run time through specific MQTT messages transmitted by a centralised trusted entity. On the same line, the work in~\cite{rausch2018emma} creates a broker network and uses an external monitoring agent to check the status of each broker. Clients are connected to brokers through local gateways: upon any change in the broker configuration (broker failure, increase in latency, etc.) the gateway reconnects client to a new broker, according to the information retrieved by the monitoring agent. Such an approach enables client mobility, dynamic broker provisioning, and broker load balancing. Finally, an example of tree-based MQTT broker topology is given in~\cite{park2018dm}, where authors propose the use of Software Define Networking (SDN) to create per-topic multicast groups in order to minimise data transfer delay. The SDN controller gathers information about clients and relative pub/sub topics from all the edge brokers through a master broker, which acts as root of the multicast tree. However, the paper assumes a static topology and no details are given on how such a root broker should be elected.

\section{Conclusion}
\label{sec:conclusion}
This paper proposes MQTT-ST, a system based on the Spanning Tree Protocol, which is able to create a distributed network of MQTT brokers. The brokers in the network generate a tree-based topology in a distributed way, which is able to fully replicate messages in every broker and to react to failures. The system has been tested in different scenarios, comparing the obtained performances with the legacy centralised solution. Future work direction will explore the integration of more complex routing strategies, besides message flooding, that can further increase the system performance. The MQTT-ST project is available for download at https://github.com/ANTLab-polimi/mosquitto.
\balance


\bibliographystyle{IEEEtran}
\bibliography{bibfile}

\end{document}